\documentclass[preprint,nofootinbib]{revtex4-1}
 \usepackage{graphicx,psfrag,epsfig}
\newcommand{\beq}{\begin{equation}}
\newcommand{\eeq}{\end{equation}}

\newcommand{\bea}{\begin{eqnarray}}
\newcommand{\eea}{\end{eqnarray}}

\newcommand{\ea}{{\it et al.}}

\begin{document}
\begin{flushright}
UMD-PP-11-011\\
July 2011\
\end{flushright}
\vspace{0.3in}
%\begin{center}
\title {\LARGE Neutrino Mixings in $SO(10)$ with Type II Seesaw and $\theta_{13}$}
%\vspace{0.5in}
\author {\bf  P. S. Bhupal Dev, R. N. Mohapatra and Matt Severson\\}
%\vspace{0.2in}
\affiliation{ Maryland Center for Fundamental Physics and
Department of Physics, University of Maryland, College Park, MD
20742, USA}
\vspace{0.5in}
%%%%%%%%%%%%%%%%%%%%%%%%%%%%%%%%%%%%%%%%%%%%%%%%%%%%%%%%%%%%%%%%%%%%%
\begin{abstract}
We analyze a class of supersymmetric $SO(10)$ grand unified theories with type II seesaw for neutrino masses,
where the contribution to PMNS matrix from the neutrino sector has an exact tri-bi-maximal (TBM) form,  dictated by a broken $S_4$ symmetry. The Higgs fields that determine the fermion masses are two {\bf 10} fields and one {\bf 126} field, with the latter simultaneously contributing to neutrino as well as charged fermion masses.  Fitting charged fermion masses and the CKM mixings lead to corrections to the TBM mixing that determine the final PMNS matrix with the predictions $\theta_{13}\simeq 4^\circ-6^\circ$ and the Dirac $C\!P$ phase to be between $-10^\circ$ and $+15^\circ$. We also show correlations between various mixing angles which can be used to test the model. 
\end{abstract}
\maketitle

%%%%%%%%%%%%%%%%%%%%%%%%%%%%%%%%%%%%%%%%%
\section{Introduction}
%%%%%%%%%%%%%%%%%%%%%%%%%%%%%%%%%%%%%%%%%
Understanding neutrino masses and mixings is an integral part 
of our attempts to unravel the flavor puzzle in particle physics. During the past decade,
 the large amount of information on neutrino masses and mixings  gained from the study of
  accelerator, reactor, solar, and cosmic ray neutrino observations has given a strong forward momentum to this journey. Several crucial pieces of the puzzle must still be found before we can begin to have a complete picture at hand; among them are the nature of the neutrino masses (Dirac vs Majorana), the mass hierarchy (normal vs inverted), the mixing angle $\theta_{13}$, and the $C\!P$ phases. 
 
 Of the large number of new experiments that are under way to answer these questions, the T2K experiment has recently announced a possible indication of
a non-zero value for $\theta_{13}$~\cite{T2K}, which has caused a great deal of excitement in the field.  The T2K lower limit, if correct,  is not far below the current experimental bound from the CHOOZ experiment~\cite{chooz} and has important theoretical implications. The MINOS experiment has also seen an excess of electron events which could be indicating a non-zero $\theta_{13}$~\cite{MINOS}, and their allowed range for $\theta_{13}$ overlaps with the T2K one. 
There have also been analyses of existing oscillation data suggesting a non-zero $\theta_{13}$~\cite{lisi}. 
Additionally, other experiments are currently searching for this important parameter~\cite{DC}, and several recent papers have attempted to 
explain the T2K values within different models~\cite{recent}; there is hope the situation will become much clearer in near future. 

A non-zero value for $\theta_{13}$ has profound implications for our understanding of the physics of neutrino mass. It is, for example, well known that maximal atmospheric neutrino mixing ($\tan \theta_{23}=1$) suggests an underlying discrete $\mu-\tau$ symmetry (denoted by $Z_{2,\mu-\tau}$) in the neutrino mass matrix, which, when  exact, leads to vanishing $\theta_{13}$~\cite{mutau} .  Depending how this symmetry is broken (e.g. in the $\mu$-sector or $e$-sector), the resulting value of $\theta_{13}$ can either be very small or not so small.
The neutrino mass matrix has also been suspected to have a larger symmetry beyond this from the observation that the  current values of the solar mixing angle seems to have a geometric value ($\tan \theta_{12}=1/\sqrt{2}$). The resulting  lepton mixing (PMNS) matrix is known as the tri-bi-maximal mixing matrix~\cite{tbm} (TBM for short). This symmetry is often denoted by  two $Z_2$ symmetries $Z_{2,S}\times Z_{2,\mu-\tau}$~\cite{lam}.  This full symmetry leads to 
zero $\theta_{13}$ and restricts the form of the neutrino mass matrix (to be called TBM matrix) to
\begin{eqnarray}
{\cal M}_\nu~=~\left(\begin{array}{ccc} b & c & c \\ c & a+b & c-a \\ c & c-a & a+b\end{array}\right),
	\label{eq:1}
\end{eqnarray}
 which is given by only three parameters.  In fact, in the above matrix, one could set $b=0$, without changing the TBM  PMNS matrix. It only affects the masses of the neutrinos. This matrix is very different from the known mass matrices in the quark sector and could be a possible clue to a unified understanding of the quark-lepton flavor puzzle. A non-zero $\theta_{13}$ suggests that the TBM PMNS mixing is not precisely the right form, and that ``large''  corrections to both the $\mu-\tau$ symmetry and TBM matrix must be present; these factors could eventually guide us towards a complete determination of the neutrino mass matrix.  Once this is accomplished, we will have passed  a major milestone in uncovering the physics of  neutrino  mass and possible underlying symmetries of the lepton sector. Of course, if observations require that the corrections to TBM mass matrix are ``large'', it would not be too implausible that the symmetries described above may only be illusory and some other mechanisms may be at work.

To explore what  other scenarios could lead us to the desired neutrino mixings with a ``large'' $\theta_{13}$,  recall that there is a large class of predictive $SO(10)$ grand unified models~\cite{so10,goran,goh,theory} in which neutrino masses arise out of a type II seesaw~\cite{type2} mechanism. These models  
provide a natural way to understand a large atmospheric mixing angle not from some symmetry, but rather from the dynamical property that in grand unified theories, the bottom and tau masses become nearly equal at GUT scale~\cite{goran}. When these models are analyzed for the full three generation case, one finds, in addition to a large $\theta_{12}$, that $\theta_{13}$ is also generally ``large''~\cite{goh}. Though the first of these results were obtained without quark $C\!P$ violation, these models have since been studied in much greater detail and including the phase. These full $C\!P$-violating models do confirm the above results including a ``large'' $\theta_{13}$ as well~\cite{so10type2}, but at the cost of severely restricting the parameter space. It turns out, however, that a slight extension of the Higgs sector by the addition of a {\bf 120} Higgs multiplet~\cite{so10120} considerably broadens the parameter space while still preserving the  ``large'' $\theta_{13}$ prediction. 

Recently, an interesting connection to the standard TBM model discussion has been noted: the type II seesaw formula for neutrino masses allows a TBM form for the neutrino mass matrix by simply a choice of fermion basis, with no additional symmetries~\cite{alta}; corrections to the TBM form then arise from the form of the charged lepton matrix, which, in our case, is determined by the $SO(10)$ constraints from quark masses and mixings.\footnote{For charged lepton corrections to tri-bi-maximal mixing outside the framework of GUT theories, see Ref.~\cite{werner}.} Strictly speaking, no bottom-tau unification is invoked in this approach. Detailed numerical analyses of these models have been carried out and  lead to excellent fits for models with {\bf 10}, {\bf 126} and {\bf 120} Higgs fields~\cite{alta,joshi}, and, yet again, a large $\theta_{13}$ is predicted. One could therefore construe the ``large''  $\theta_{13}$ prediction of these models as an indication of grand unified origin of neutrino masses (especially of the kind noted), which was anyway suspected as a possibility due to the near-GUT seesaw scale.

As mentioned above, the near-tri-bi-maximal PMNS form in this class of GUT theories is related to the dynamics of the model rather than to any symmetry. Of course, to understand the particular Yukawa textures, one may need to invoke some symmetries, but still those symmetries are not directly related to the $\theta_{13}$ value. We are therefore faced with two contrasting but attractive approaches to current neutrino observations: one based on leptonic symmetries, and another based on grand unification hypothesis. It is clearly important that more work be done to uncover which is the path chosen by nature; in this paper, we further investigate the grand unification approach.

One straightforward way to establish  that a ``large'' $\theta_{13}$ is a generic prediction of $SO(10)$ models with type II seesaw and their associated dynamical properties, rather than a symmetry, is to study more of such models and establish their predictions. A particularly simple class of models are defined by the minimal choice of Higgs fields {\bf 10} and {\bf 126}, together with either a {\bf 120}~\cite{so10120} (as already noted) or an extra {\bf 10}~\cite{dmm} contributing to fermion masses. The latter class of models, to the best of our knowledge, has not been thoroughly scrutinized numerically. In this paper, we focus on them, since, as has been recently pointed out~\cite{dmm}, they seem to give qualitatively the right picture for not just neutrino masses but quarks as well. 
%An interesting property of this class of models is that by an appropriate choice of the basis for matter fields in $SO(10)$, one can cast the neutrino mass matrix in the TBM form~\cite{alta}, so that any correction to the TBM form of the PMNS mixing must arise from the charged lepton sector. That, in turn, is related to the quark masses and mixings, establishing an intimate connection between the quark and the lepton sector.
 It was shown in Ref.~\cite{dmm} that reasonably well-satisfied versions of the GUT scale relations $m_b\simeq m_\tau$ and $m_\mu \simeq -3m_s$ emerge out of an $S_4$ flavor symmetry in an $SO(10)$ GUT model of the above type.  It was also noted in this paper that the TBM form of the neutrino mass matrix is  dictated  by the $S_4$ symmetry breaking. From our quantitative analysis of this model, we first find that the Yukawa texture predicted by the minimal version 
 of the model~\cite{dmm} needs to be supplemented by additional effective  GUT scale Yukawa couplings in order to come close to observations. The improved model has only twelve parameters and is therefore predictive in the neutrino sector. We find that the model leads to a prediction for 
 $\theta_{13}\sim 4^\circ-6^\circ$ and Dirac $C\!P$ phase between $-10^\circ$ and $+15^\circ$; this value of $\theta_{13}$  supports the generic expectation for this class of theories, as was anticipated above. We also argue that there is a definite kind of correlation between the $\theta_{13}$ and $\theta_{23}$, which can be different from non-GUT symmetry-based approaches to  $\theta_{13}$. It is also interesting that this value is consistent with the recent T2K range for this parameter.
%%%%%%%%%%%%%%%%%%%%%%%%%%%%%%%%%%%%%%%%%%%%%%%%%%%%%%%%%%%%
%%%%%%%%%%%%%%%%%%%%%%%%%%%%%%%%%%%%%%%%%%%%%%%%%%%%%%%%%%%%
\section{ Details of the model}
The class  of $SO(10)$ models in which we are interested have two {\bf 10} Higgs fields (denoted by $H,H'$) and a pair of ${\bf 126}+\overline{\bf 126}$ (denoted by $\Delta+\bar{\Delta}$). The $SO(10)$ invariant Yukawa couplings of the model are given by:
\begin{eqnarray}
{\cal L}_Y~=~h_0\psi\psi H~+~h'_0\psi\psi H'~+~f_0\psi\psi\bar{\Delta}
\end{eqnarray}
where $\psi$'s denote the {\bf 16} dimensional  spinors of $SO(10)$ which contain all the matter fields of each generation, and so there are three such fields, though we have suppressed the generation indices. The Yukawa couplings are $3\times 3$ matrices in generation space.
The effective Yukawa couplings $f_0, h_0, h'_0$ are assumed to have descended from a higher scale theory which has an $S_4$ symmetry broken by flavon fields with particularly aligned vacuum expectation values (vevs) (see e.g. Ref.~\cite{dmm}). We do not need to know the detailed form of these flavon interactions for our analysis in this paper, and we will simply write down the effective form of the $h,h',f$ that follow from it. Before doing that, we wish to point out that it is the $f_0$ coupling which is responsible for neutrino masses via type II seesaw mechanism, and it also contributes to charged fermion masses. We can choose it to give the tri-bi-maximal form for ${\cal M}_\nu$ either by a choice of the basis of matter fields~\cite{dmm,alta}  or by the breaking of the $S_4$ symmetry~\cite{dmm} or other symmetry group~\cite{king}.
This puts the neutrino mass matrix in a form that, upon diagonalization, leads to tri-bi-maximal mixing prior to charged lepton corrections. This requires that we have the $f_0$ coupling in the form:
\begin{eqnarray}
f_0 \propto {\cal M}_\nu= \kappa \left(\begin{array}{ccc}0 & m_1 & m_1 \\
m_1 & m_0 & m_1-m_0\\
m_1 & m_1-m_0 & m_0
\end{array}\right),
\label{eq:3}
\end{eqnarray}
where we have used Eq.~(\ref{eq:1}) with rescaled variables $a = \kappa m_0$ and $c = \kappa m_1$. Note that as in Ref.~\cite{dmm}, we have taken $b=0$ in 
Eq.~(\ref{eq:1}).
%this corresponds to $b=0$ in Eq.~(\ref{eq:1}), 
%which is dictated by the $S_4$ symmetry in this 
%model~\cite{dmm}. 
The proportionality constant between $f_0$ and ${\cal M}_\nu$ is determined by the left triplet vev in {\bf 126} responsible for type II seesaw.  Note that $U_{PMNS}=V^\dagger_{\ell}V_\nu$ (where $V_{\ell}$ and $V_{\nu}$ are the unitary matrices that diagonalize the charged lepton and neutrino mass matrices 
respectively) so that we will necessarily have corrections to the TBM mixing coming from the charged lepton mass matrix.  Note further that since the $f_0$ matrix also contributes to the quark and charged lepton masses, neutrino masses and quark masses are connected, making the model predictive.
The formulae for the quark and charged lepton masses in this model are given by:
\begin{eqnarray}
M_u &= & h~+~r_2 f~+~r_3h', \nonumber \\
M_d &=& \frac{r_1}{\tan\beta}(h+f+h'),\label{eq:4} \\
M_\ell &=& \frac{r_1}{\tan\beta}(h-3f+h'), \nonumber
\end{eqnarray}
where $f,h,h'$ are related to $f_0, h_0, h'_0$ through Higgs vevs~\cite{dmm}. 
In Ref.~\cite{dmm}, the $S_4$ symmetry constrains $h$ to be a rank one matrix of the form: 
\begin{equation}
	h = \left(\begin{array}{ccc} 0 & 0 & 0 \\ 0 & 0 & 0 \\0 & 0 & M\end{array}\right),
	\end{equation}
and the form of $h'$ to be
\begin{equation}
	h' = \left(\begin{array}{ccc} 0 & \delta & -\delta \\ \delta & 0 & 0 \\-\delta & 0 & 0
\end{array}\right).
\end{equation}
The parameters $m_0, m_1$ in Eq.~(\ref{eq:3}) are chosen to be real. The parameters $r_{1,2,3}$ in Eqs.~(\ref{eq:4}) represent the ratio of the different standard model (SM) doublet vevs in the theory. There are three SM doublets and hence six vevs; three of them are absorbed to redefine  the Yukawa couplings from dimensionless $h_0, h'_0, f_0$ to $h,h',f$ with dimensions of mass. Since we have chosen $h$ to be in the form above due to the $S_4$ symmetry breaking, it has only one parameter. The $f$ matrix has two real parameters,  and $h'$ has only one parameter, which can be chosen to be complex, for a total of eight parameters in the charged fermion sector. While this model has a number of attractive features as noted in Ref.~\cite{dmm}, it fails to reproduce some details of the quark mixings, e.g. both $V_{cb}$ and $V_{ub}$ come out to be much too small compared to their extrapolated values at the GUT scale for all $\tan\beta$; the CKM phase also comes out too small. We therefore amend this model by adding extra structure to the $h'$ matrix while keeping all other couplings as they were.  We choose $h'$ to have the form:  
\begin{eqnarray}
h' = \left(\begin{array}{ccc} \delta^{\prime\prime} & \delta &-\delta+ \delta^\prime \\ \delta & 0 & d
\\ -\delta+\delta^\prime & d & 0\end{array}\right),
\end{eqnarray}
which can be generated by choice of flavon fields and the alignment of their vevs. 
%We do not discuss this process further here.
The neutrino mass matrix is unaffected by this addition, but the quark
and charged lepton mass matrices are now
\begin{eqnarray}
M_\ell &=& \frac{r_1}{\tan\beta}\left( \begin{array}{ccc}
\delta^{\prime \prime} & -3m_1+\delta & -3m_1-\delta+\delta^\prime \\
-3m_1+\delta & -3m_0 & 3m_0-3m_1+d \\
-3m_1-\delta+\delta^\prime & 3m_0-3m_1+d & -3m_0+M \end{array} \right) \nonumber \\
 M_d &=& \frac{r_1}{\tan\beta}\left( \begin{array}{ccc}
\delta^{\prime \prime} & m_1+\delta & m_1-\delta+\delta^\prime \\
m_1+\delta & m_0 & -m_0+m_1+d \\
m_1-\delta+\delta^\prime & -m_0+m_1+d & m_0+M \end{array} \right) \label{eq:8}
\\
 M_u &=& \left( \begin{array}{ccc}
r_3 \delta^{\prime \prime} & r_2 m_1+r_3 \delta & r_2 m_1-r_3 \left(\delta-\delta^\prime \right) \\
r_2 m_1+r_3 \delta & r_2 m_0 & -r_2 m_0+r_3 \left(m_1+d \right) \\
r_2 m_1-r_3 \left(\delta-\delta^\prime \right) & -r_2 m_0+r_3 \left(m_1+d \right) & r_2 m_0+M \end{array} \right) \nonumber
\end{eqnarray}
%%%%%%%%%%%%%%%%%%%%%%%%%%%%%%%%%%%%%
%%%%%%%%%%%%%%%%%%%%%%%%%%%%%%%%%%%%%
\section{Predictions of the model}
The model has eleven parameters if we choose all except $\delta'$ real (twelve parameters when we allow $\delta$ complex to study the allowed range of 
Dirac $C\!P$ phase). Recall that the model with {\bf 10}, {\bf 126} and {\bf 120} has a total of seventeen parameters~\cite{alta}. In that sense ours is a more economical one and is quite predictive.
%%%%%%%%%%%%%%%%%%%%%%%%%%
Before proceeding with the numerical analysis discussion, we note a few results that can be derived analytically if we assume the hierarchy $M \gg m_0, d \gg m_1, \delta, \delta^\prime, \delta^{\prime\prime}$:
\begin{eqnarray} 
\frac{r_1}{\tan\beta} &\simeq& \frac{m_b}{m_t}; ~ ~ M \simeq m_t;
\nonumber \\
m_c \simeq r_2 m_0, ~ ~ r_2 &\simeq& \frac{m_b m_c}{m_t m_s} \quad \Rightarrow \quad m_0 \simeq \frac{m_t m_s}{m_b}.
\end{eqnarray}
Diagonalizing the matrices in Eqs.~(\ref{eq:8}) gives the charged fermion masses, and the combination $V_u^\dagger V_d$ (where $V_u$ and $V_d$ diagonalize 
the up- and down-sector quark masses respectively) gives the CKM matrix. Approximate expressions for the mass eigenvalues and the CKM mixing matrices 
are given by
\begin{eqnarray}
M_\ell^D &\simeq& \frac{r_1}{\tan\beta}\left( \begin{array}{ccc}
\delta^{\prime \prime}+\frac{(-3m_1+\delta)^2}{\delta^{\prime \prime}+3m_0} & & \\
 & -3m_0-\frac{(-3m_1+\delta)^2}{3m_0+\delta^{\prime \prime}}-\frac{[3(m_0-m_1)+d]^2}{M} & \\
 & & -3m_0+M \end{array} \right) \nonumber \\
M_d^D &\simeq& \frac{r_1}{\tan\beta}\left( \begin{array}{ccc}
\delta^{\prime \prime}+\frac{(m_1+\delta)^2}{\delta^{\prime \prime}-m_0} & & \\
 & m_0+\frac{(m_1+\delta)^2}{m_0-\delta^{\prime \prime}}-\frac{(-m_0+m_1+d)^2}{M} & \\
 & & m_0+M \end{array} \right) \\
M_u^D &\simeq& \left( \begin{array}{ccc}
r_3 \delta^{\prime \prime}+\frac{(r_2 m_1+r_3 \delta)^2}{r_3 \delta^{\prime \prime}-r_2 m_0} & & \\
 & r_2 m_0+\frac{(r_2 m_1+r_3 \delta)^2}{r_2 m_0-r_3 \delta^{\prime \prime}}-\frac{[-r_2 m_0+r_3(m_1+d)]^2}{M} & \\
 & & r_2 m_0+M \end{array} \right) \nonumber
 \end{eqnarray}
and
\begin{eqnarray}
V_d &\simeq& \left( \begin{array}{ccc}
1 & \frac{m_1+\delta}{m_0-\delta^{\prime \prime}} & \frac{m_1-\delta+\delta^\prime}{-\delta^{\prime \prime}+m_0+M} \\
\frac{m_1+\delta}{\delta^{\prime \prime}-m_0} & 1 & \frac{-m_0+m_1+d}{M} \\
\frac{m_1-\delta+\delta^\prime}{\delta^{\prime \prime}-m_0-M} & \frac{m_0-m_1-d}{M} & 1 \end{array} \right) \nonumber \\
V_u &\simeq& \left( \begin{array}{ccc}
1 & \frac{r_2 m_1+r_3 \delta}{r_2 m_0-r_3 \delta^{\prime \prime}} & \frac{r_2 m_1-r_3 (\delta-\delta^\prime)}{-r_3 \delta^{\prime \prime}+r_2 m_0+M} \\
\frac{r_2 m_1+r_3 \delta}{r_3 \delta^{\prime \prime}-r_2 m_0} & 1 & \frac{-r_2 m_0+r_3 (m_1+d)}{M} \\
\frac{r_2 m_1-r_3 (\delta-\delta^\prime)}{r_3 \delta^{\prime \prime}-r_2 m_0-M} & \frac{r_2 m_0-r_3 (m_1+d)}{M} & 1 \end{array} \right)
\end{eqnarray}
Additionally, note that the resulting corresponding expression for the Cabibbo angle is
\begin{equation}
V_{us} \simeq \frac{m_1+\delta}{m_0-\delta^{\prime \prime}}-\frac{r_2 m_1+r_3 \delta}{r_2 m_0-r_3 \delta^{\prime \prime}}
\end{equation} 

Using a sufficient set of the individual expressions for $V_{ij}$ and $m_f$ above, as well as the ratio $m_1/m_0$ from the neutrino sector (to be 
discussed later), we solve a system of equations against experimental values for the charged fermion masses and quark mixings to find an analytical solution with approximate values for the input parameters; this solution is then used to generate predictions for the neutrino sector and is made statistically robust through numerical analysis. A best fit value for the input parameters is given in Table~\ref{table:1}, and the resulting mass and mixing parameter values  are given in Table~\ref{table:2}. 
\begin{table}[h!]
\begin{center}
\begin{tabular}{||c|c|c||}	\hline\hline
	& $\tan\beta = 10$ & $\tan\beta = 55$ \\
	\hline
	$M$ (GeV) & 88.2 & 106.2\\
	$m_0$ (GeV) & 1.435 & 1.382\\
	$m_1$ (GeV) & 0.275 & 0.275\\
	$\delta$ (GeV) & 0.2850 & 0.2605\\
	$\delta^\prime$ (GeV) & 0.463 - 0.279$\imath$ & 0.529 - 0.335$\imath$\\
	$\delta^{\prime \prime}$ (GeV) & -0.0652 & -0.0767\\
	$d$ (GeV) & 3.78 & 4.31\\
	$r_1/ \tan\beta$ & 0.0153 & 0.0159\\
	$r_2$ & 0.130 & 0.129\\
	$r_3$ & -0.06 & -0.07\\
	\hline\hline
\end{tabular}
\caption{Best fit values for the model parameters. Note that adding a small imaginary part to $\delta$ will give us a 
non-negligible Dirac $C\!P$ phase, as shown later in Figure~\ref{fig:4}.}
\label{table:1}
 \end{center}
 \end{table}
 Note that while we get a higher value for $m_b$ and slightly lower values for $m_s$ and $m_c$,  all are within reasonable statistical deviation from 
 extrapolated values in the literature~\cite{das}. Similarly, our predictions for $m_d$ and $m_u$ are somewhat higher than those obtained in Ref.~\cite{das}, but we believe there could easily be  instanton corrections to the light quark masses, which could change these extrapolated values. It is nevertheless remarkable that we are able to reproduce all other parameters in the charged fermion sector so well.
 \begin{table}[h!]
	 \begin{center}
		 \begin{tabular}{||c|c|c|c|c|}\hline\hline
			 & \multicolumn{2}{c|}{$\tan\beta=10$} & 
			 \multicolumn{2}{c|}{$\tan\beta=55$} \\ \cline{2-5}
			 & best fit & RG extrapolated & best fit & RG extrapolated\\
			 \hline\hline
			 $m_e$ (MeV) & 0.3587 & $0.3585^{+0.0003}_{-0.0003}$
			 & 0.3563 & $0.3565^{+0.0002}_{-0.0010}$\\
			 $m_\mu$ (MeV) & 75.6865 & $75.6715^{+0.0578}_{-0.0501}
			 $ & 75.3359 & $75.2938^{+0.0515}_{-0.1912}$\\
			 $m_\tau$ (GeV) & 1.2927 & $1.2922^{+0.0013}_{-0.0012}$
			 & 1.6272 & $1.6292^{+0.0443}_{-0.0294}$\\
			 $m_d$ (MeV) & 3.8587 & $1.5036^{+0.4235}_{-0.2304}$ & 
			 4.0202 &
			 $1.4967^{+0.4157}_{-0.2278}$\\
			 $m_s$ (MeV) & 23.6026 & $29.9454^{+4.3001}_{-4.5444}$ & 
			 23.1619 & $29.8135^{+4.1795}_{-4.4967}$\\ 
			 $m_b$ (GeV) & 1.3726 & $1.0636 ^{+0.1414}_{-0.0865}$ & 
			 1.7078 &
			 $1.4167^{+0.4803}_{-0.1944}$\\
			 $m_u$ (MeV) & 1.9772 & $0.7238^{+0.1365}_{-0.1467}$ & 
			 3.6311 &$0.7244^{+0.1219}_{-0.1466}$\\ 
	$m_c$ (MeV) & 177.3862 & $210.3273^{+19.0036}_{-21.2264}$ & 177.6719 & 
	$210.5049^{+15.1077}_{-21.1538}$\\
	$m_t$ (GeV) & 88.3886 & $82.4333^{+30.2676}_{-14.7686}$ & 106.3806 &
	$95.1486^{+69.2836}_{-20.6590}$\\
	$V_{us}$ & 0.2230 &$0.2243\pm 0.0016$ & 0.2233 & $0.2243\pm 0.0016$\\
	$V_{ub}$ & 0.0032 & $0.0032\pm 0.0005$ & 0.0032 & $0.0032\pm 0.0005$\\
	$V_{cb}$ & 0.0349 & $0.0351\pm 0.0013$ & 0.0352 & $0.0351\pm 0.0013$\\
	$\delta_{\rm CKM}$ & $-64.35^\circ$ & $(-60\pm 14)^\circ$ & $-61.84^\circ$ &$ (-60\pm 14)^\circ$\\
	\hline\hline
\end{tabular}
\caption{The best fit values of the quark and charged lepton masses and 
the most relevant quark mixing parameters. The 
$1\sigma$ experimental values extrapolated by MSSM renormalization group (RG) equations to the GUT scale~\cite{das} are also shown 
for comparison.}
\label{table:2}
\end{center}
\end{table}

For the neutrino sector, the structure of the mass matrix in the model of Ref.~\cite{dmm} and in this amended model is given by
\begin{equation}
M_\nu = \kappa \left(\begin{array}{ccc}
0 & m_1 & m_1\\
m_1 & m_0 & m_1-m_0\\
m_1 & m_1-m_0 & m_0\cos^2\alpha \end{array}\right)
\end{equation}
where $\kappa$ is a scaling factor determined from experimental data, and $\alpha$ is the mixing angle for the third generation matter fermion $\psi$ with the vector-like field $\psi_V$ specific to the model. The limit $\alpha=0$ gives the strict TBM form for the neutrino mass matrix given in Eq.~(\ref{eq:3}) 
when the mass eigenvalues are 
\begin{equation}
	m_{\nu_1} = -\kappa m_1, \quad m_{\nu_2} = 2\kappa m_1, \quad m_{\nu_3} = \kappa~(2m_0-m_1)
\end{equation}
and the solar-to-atmospheric mass-squared ratio is given by
\begin{equation}
	\frac{\Delta m^2_\odot}{\Delta m^2_{\rm atm}} = \frac{3}{4}
	\frac{\lambda^2}{1-\lambda} 
	\label{eq:15}
\end{equation}
where $\lambda\equiv m_1/m_0$. To fit the experimental data,  $\Delta m^2_\odot / \Delta m^2_{\rm atm} \sim 0.03$, which corresponds to $\lambda \sim 0.2$ from Eq.~(\ref{eq:15}). This constraint can be relaxed for the case of $\alpha\neq 0$, which is already required for the top Yukawa coupling ($\propto \sin^2\alpha$) to be non-zero~\cite{dmm}, though we do not have much freedom for the value of $\lambda$ anyway, as it is tightly constrained by the quark sector. Numerically, we find that the allowed range of $\alpha$ is $5^\circ-25^\circ$ in order to fit the observed neutrino data.

Noting the charged lepton rotation matrix from the ansatz given by Eq.~(\ref{eq:8}):
\begin{eqnarray}
V_\ell &\simeq& \left( \begin{array}{ccc}
1 & \frac{3m_1-\delta}{\delta^{\prime \prime}+3m_0} & \frac{-3m_1-\delta+\delta^\prime}{-\delta^{\prime \prime}-3m_0+M} \\
\frac{-3m_1+\delta}{\delta^{\prime \prime}+3m_0} & 1 & \frac{3(m_0-m_1)+d}{M} \\
\frac{-3m_1-\delta+\delta^\prime}{\delta^{\prime \prime}+3m_0-M} & \frac{3(m_1-m_0)-d}{M} & 1 \end{array} \right),
\end{eqnarray}
and given the TBM form of the matrix that diagonalizing the neutrino mass 
matrix:
\begin{eqnarray}
	V_{TMB} = \left(\begin{array}{ccc}
		\sqrt{\frac{2}{3}} & \sqrt{\frac{1}{3}} & 0\\
		-\sqrt{\frac{1}{6}} & \sqrt{\frac{1}{3}} & -\sqrt{\frac{1}{2}}\\
		-\sqrt{\frac{1}{6}} & \sqrt{\frac{1}{3}} & \sqrt{\frac{1}{2}}
	\end{array}\right),
\end{eqnarray}
we can write an approximate analytical form of the PMNS neutrino mixing matrix: 
\begin{eqnarray}
	&& U_{PMNS} = V_\ell^\dagger V_{TBM} \simeq V_{TBM}+ \nonumber \\
&& \left(\begin{array}{ccc}
	\frac{1}{\sqrt 6}\left(\frac{3m_1-\delta}{\delta^{\prime \prime}+3m_0}-\frac{3m_1+\delta-\delta^{\prime^\ast}}{M-3m_0-\delta^{\prime \prime}}
	\right)
	& \frac{1}{\sqrt{3}}\left(\frac{-3m_1+\delta}{\delta^{\prime \prime}+3m_0}+\frac{3m_1+\delta-\delta^{\prime^\ast}}{M-3m_0-\delta^{\prime \prime}}\right)
	& \frac{1}{\sqrt{2}}\left(\frac{3m_1-\delta}{\delta^{\prime \prime}+3m_0}+\frac{3m_1+\delta-\delta^{\prime^\ast}}{M-3m_0-\delta^{\prime \prime}}\right) \\
	\frac{1}{\sqrt 6}\left(\frac{3(m_0-m_1)+d}{M}+\frac{2(3m_1-\delta)}{\delta^{\prime \prime}+3m_0}\right) 
	& \frac{1}{\sqrt{3}}\left(\frac{3(m_1-m_0)-d}{M}+\frac{3m_1-\delta}{\delta^{\prime \prime}+3m_0}\right) 
	& \frac{3(m_1-m_0)-d}{\sqrt{2}M} \\
	\frac{1}{\sqrt 6}\left(\frac{3(m_1-m_0)-d}{M}-\frac{2(3m_1+\delta-\delta^{\prime^\ast})}{M-3m_0-\delta^{\prime \prime}}\right) 
	& \frac{1}{\sqrt{3}}\left(\frac{3(m_0-m_1)+d}{M}-\frac{3m_1+\delta-\delta^{\prime^\ast}}{M-3m_0-\delta^{\prime \prime}}\right) 
	& \frac{3(m_1-m_0)-d}{\sqrt{2}M} \end{array} \right) \nonumber \\
\label{eq:17}
\end{eqnarray}

\begin{figure}[!htb]
\centering
\includegraphics[width=10.0cm]{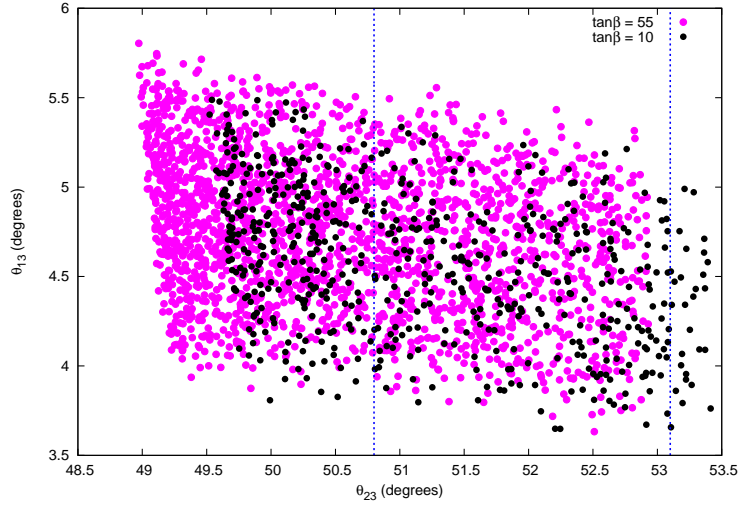}
\caption{Correlation between $\theta_{13}$ and $\theta_{23}$ predicted in our model satisfying all the charged fermion sector constraints. The thin (thick) 
dotted vertical line 
is the current $2\sigma$ ($3\sigma$) upper limit for the atmospheric mixing angle.}
\label{fig:1}
\end{figure}

\begin{figure}[!htb]
\centering
\includegraphics[width=10.0cm]{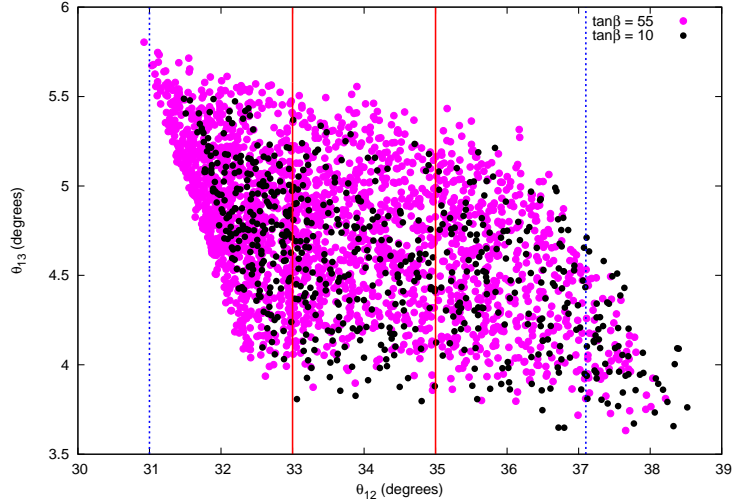}
\caption{Correlation between $\theta_{13}$ and $\theta_{12}$ predicted in our model. The solid and dotted vertical lines are the current 
$1\sigma$ and $3\sigma$ limits respectively for the solar mixing angle.}
\label{fig:2}
\end{figure}
\begin{figure}[!htb]
\centering
\includegraphics[width=10.0cm]{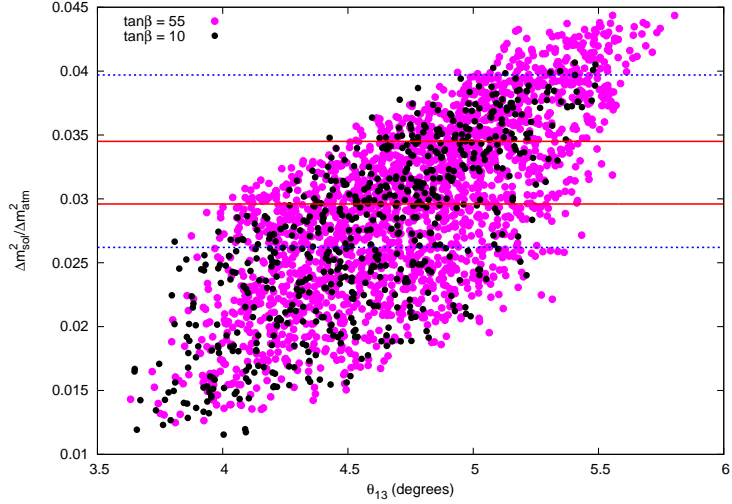}
\caption{Correlation between $\Delta m^2_\odot / \Delta m^2_{\rm atm}$ and 
$\theta_{13}$ predicted in our model. The solid and dotted horizontal lines are the current 
$1\sigma$ and $3\sigma$ limits respectively for the solar-to-atmospheric mass squared ratio.}
\label{fig:3}
\end{figure}
%{\bf Need value for c to give the neutrino mass eigenvalues here.}\\

\begin{figure}[!htb]
\centering
\includegraphics[width=10.0cm]{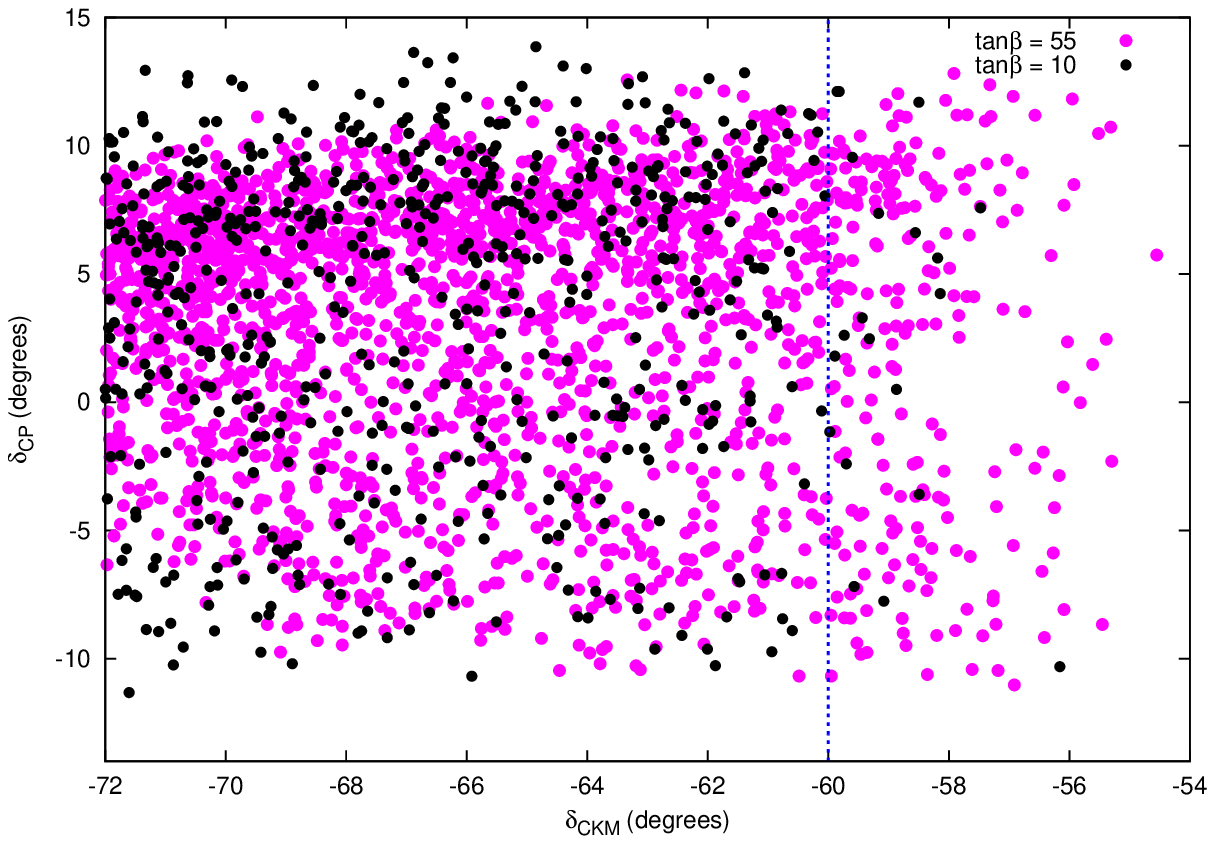}
\caption{Correlation between the Dirac and CKM $C\!P$ phase. The dotted vertical line is the central value of the observed CKM $C\!P$ phase. The spread in phase in the model arises from complexifying the parameter $\delta$ so that it 
does not pull the CKM phase out of its allowed range.}
\label{fig:4}
\end{figure}

The exact numerical results  for neutrino mixing corresponding to the quark sector fit from above are given in Figures \ref{fig:1}-\ref{fig:4}. Figures \ref{fig:1} and \ref{fig:2} show the relationships between $\left(\theta_{23}, \theta_{13}\right)$ and $\left(\theta_{12}, \theta_{13}\right)$, respectively. Note that the value of $\theta_{13}$ is large, though not so large as the $6^\circ$-$8^\circ$ central value of T2K result gives. Also 
note that the atmospheric mixing angle $\theta_{23}$ is always larger than the maximal value of $45^\circ$; this agrees with the analytical form given in 
Eq.~(\ref{eq:17}) in which the corrections from the charged lepton sector are always positive. Figure \ref{fig:3} shows the correlation between 
$\left(\theta_{13},\Delta m^2_{\odot}/\Delta m^2_{\rm atm}\right)$. The solid and dotted lines are the current $1\sigma$ and $3\sigma$ limits for 
the best fit values of the observed neutrino oscillation parameters~\cite{lisinew} (using the new reactor neutrino fluxes). Figure \ref{fig:4} shows the correlation between the Dirac and CKM $C\!P$ phases. These correlations between different mixing parameters could be used to test the model once the current uncertainties in both $\theta_{12}$ and $\theta_{23}$ are reduced and a more precise value for $\theta_{13}$ has been 
determined. Our model also predicts small Majorana phases ($\sim 1^\circ$). 

\section{Summary}
We have analyzed the predictions of an $SO(10)$ model with type II 
seesaw for neutrino masses and Yukawa couplings involving two {\bf 10} Higgs fields and 
one {\bf 126} Higgs field, with all the couplings derivable from a broken $S_4$ 
symmetry. The model has at most twelve parameters and is thus a relatively economical one when 
compared to other models discussed in the literature. It gives a fairly good 
fit to the charged fermion masses as well as an excellent fit to the CKM parameters, and it also predicts the neutrino mixing angles $\theta_{12}$, $\theta_{23}$ as well as 
$\Delta m^2_\odot/\Delta m^2_{atm}$ in agreement with observation. 
Furthermore, it predicts a value for $\theta_{13}$ between $4^\circ-6^\circ$, near the lower end of the current T2K allowed range. With more accurate determination of 
$\theta_{13}$ -- especially its correlation with $\theta_{23}$, which our model predicts to be strictly larger than 45$^\circ$ -- the model could be tested.
Finally, the model predicts a normal hierarchy for the neutrinos and hence an effective neutrino mass in neutrino-less double beta decay, which is a few milli-electron-volts and is thus not observable in the current round of the searches for this process.
%%%%%%%%%%%%%%%%%%%%%%%
\section{Acknowledgment}
The present work is supported by the National Science Foundation grant number PHY-0968854. We thank M. K. Parida for useful comments and suggestions. 
%%%%%%%%%%%%%%%%%%%%%%%%%%%%%%%%%%

\end{document}